\documentclass[8.5pt,twoside,twocolumn]{article}
\usepackage[super,sort&compress,comma]{natbib}
\usepackage{mhchem}
\usepackage{times,mathptmx}
\usepackage{sectsty}
\usepackage{balance}

\usepackage{graphicx} 
\usepackage{lastpage}
\usepackage[format=plain,justification=raggedright,singlelinecheck=false,font=small,labelfont=bf,labelsep=space]{caption}
\usepackage{fancyhdr}
\usepackage{color}
\usepackage{extarrows}

\newcommand{\ib}[1]{{\color{black}#1}}

\newcommand*{\emphj}[1]{{\emph{#1}}}
\begin{document}

\thispagestyle{plain}
\fancypagestyle{plain}{
\renewcommand{\headrulewidth}{1pt}}
\renewcommand{\thefootnote}{\fnsymbol{footnote}}
\renewcommand\footnoterule{\vspace*{1pt}%
\hrule width 3.4in height 0.4pt \vspace*{5pt}}
\setcounter{secnumdepth}{5}

\makeatletter
\def\subsubsection{\@startsection{subsubsection}{3}{10pt}{-1.25ex plus -1ex minus -.1ex}{0ex plus 0ex}{\normalsize\bf}}
\def\paragraph{\@startsection{paragraph}{4}{10pt}{-1.25ex plus -1ex minus -.1ex}{0ex plus 0ex}{\normalsize\textit}}
\renewcommand\@biblabel[1]{#1}
\renewcommand\@makefntext[1]%
{\noindent\makebox[0pt][r]{\@thefnmark\,}#1}
\makeatother
\renewcommand{\figurename}{\small{Fig.}~}
\sectionfont{\large}
\subsectionfont{\normalsize}

\fancyfoot{}
\fancyfoot[RO]{\footnotesize{\sffamily{1--\pageref{LastPage} ~\textbar  \hspace{2pt}\thepage}}}
\fancyfoot[LE]{\footnotesize{\sffamily{\thepage~\textbar\hspace{3.45cm} 1--\pageref{LastPage}}}}
\fancyhead{}
\renewcommand{\headrulewidth}{1pt}
\renewcommand{\footrulewidth}{1pt}
\setlength{\arrayrulewidth}{1pt}
\setlength{\columnsep}{6.5mm}
\setlength\bibsep{1pt}

\newcommand{\alt}{\raisebox{-0.3ex}{$\stackrel{<}{\sim}$}}
\newcommand{\agt}{\raisebox{-0.3ex}{$\stackrel{>}{\sim}$}}

\twocolumn[
  \begin{@twocolumnfalse}
\ib{\noindent\LARGE{\textbf{An important impact of the molecule-electrode couplings asymmetry on the efficiency of bias-driven redox processes in molecular junctions}}}
\vspace{0.6cm}

\noindent\large{\textbf{Ioan B\^aldea $^{\ast}$
\textit{$^{a\dag}$}
}}\vspace{0.5cm}


\noindent \textbf{\small{Published: Phys.~Chem.~Chem.~Phys.~2015, {\bf 17}, 15756-15763; DOI: 10.1039/C5CP01805F}}
\vspace{0.6cm}

\noindent 
\normalsize{Abstract:\\
Two recent experimental 
and theoretical studies
(\emph{Proc.\ Natl.\ Acad.\ Sci.\ U.~S.~A.} {\bf 2014}, 111, 1282-1287; 
\emph{Phys.\ Chem.\ Chem.\ Phys.}\ {\bf 2014}, 16, 25942-25949) 
have addressed the problem of tuning molecular charge and vibrational properties of single molecules 
embedded in nanojunctions. These are molecular characteristics escaping so far to an efficient experimental 
control in broad ranges. 
Here, we present a general argument demonstrating why, out of various experimental platforms possible, 
those wherein active molecules are asymmetrically coupled to electrodes are to be preferred to those symmetrically 
coupled for achieving 
a(n almost) complete redox process, and why electrochemical environment has advantages over ``dry'' setups.
This study aims at helping to nanofabricate molecular junctions
using the most appropriate platforms 
enabling the broadest possible bias-driven control of the redox state and vibrational modes of single 
molecules linked to electrodes.

$ $ \\  

}
\vspace{0.5cm}
 \end{@twocolumnfalse}
  ]


\footnotetext{\textit{$^{a}$~Theoretische Chemie, Universit\"at Heidelberg, Im Neuenheimer Feld 229, D-69120 Heidelberg, Germany.}}
\footnotetext{\dag~E-mail: ioan.baldea@pci.uni-heidelberg.de.
Also at National Institute for Lasers, Plasmas, and Radiation Physics, Institute of Space Sciences,
Bucharest, Romania}
%
%
\section{Introduction}
\label{sec:intro}
In spite of impressive advances in nanoelectronics,
\cite{Molen:13,Ratner:13b,Choi:08,Loertscher:13b} 
detailed characterization and control of molecular properties under \emphj{in situ} conditions continue
to remain important challenges for fabricating and understanding single-molecule junctions.
To this aim, more recent studies have emphasized the need to go beyond electronic transport.
\cite{Tao:06e,Natelson:08c,Venkataraman:13a} 
Vibrational properties studied via 
inelastic electron tunneling spectroscopy (IETS)
\cite{Ho:98}
and surface enhanced Raman spectroscopy (SERS)
\cite{Tao:06e,Schatz:06,Natelson:08c,Liu:11,Wandlowski:11c,Konishi:13a}
may represent such valuable piece of information, ideally if they are acquired concomitantly with
transport measurements. This has been demonstrated in a recent joint experimental theoretical study 
on fullerene-based electromigrated junctions \cite{Natelson:14}
and a theoretical study \cite{Baldea:2014f} on viologen-based junctions 
in electrochemical (EC) scanning tunneling microscope (STM) setup, which was based on 
experimental data reported previously.\cite{Wandlowski:08} 

Refs.~\citenum{Natelson:14} and \citenum{Baldea:2014f}
indicated that a significant tuning on the vibrational frequencies and Raman scattering 
intensities can be obtained via 
bias-driven changes in the charge of the active molecule in a current  carrying state;
tuning the molecular charge via applied biases enables to control chemical bond strengths and, 
thence, vibrational properties.

Obviously, the broadest control that can be achieved pursuing this route 
corresponds to fully change the average redox state of the molecule by adding an entire electron.
To be specific, we limit ourselves to n-type (LUMO-mediated) conduction 
(orbital energy offset $\varepsilon_0 > 0$), 
as this is the case for the molecular junctions 
of refs.~\citenum{Natelson:14} and \citenum{Baldea:2014f}.\cite{oxidation-homo}
Reaching this ideal limit was impossible in the experiments of ref.~\citenum{Natelson:14}. 
On the other hand, ref.~\citenum{Baldea:2014f} indicated that an almost perfect reduced ($n \approx 1$) 
state can be reached in electrolytically gated junctions.\cite{Wandlowski:08}

Are the obstacles to obtaining a full reduction in 
some molecular junctions
merely of technical nature? 
This is the question that initiated the present study.
By comparing performances of various nanofabrication platforms and identifying
the ones (depicted in \figurename\ref{fig:ReductionAsymmetricPlatform}a and \figurename\ref{fig:i_n_vs_e0}a below)
that are most advantageous for achieving an  almost complete reduction,
the present study aims at helping  
to design molecular junctions enabling the broadest bias-driven control 
over molecular charge and vibrational properties.
\section{Model}
\label{sec:model}
\begin{figure}[htb]
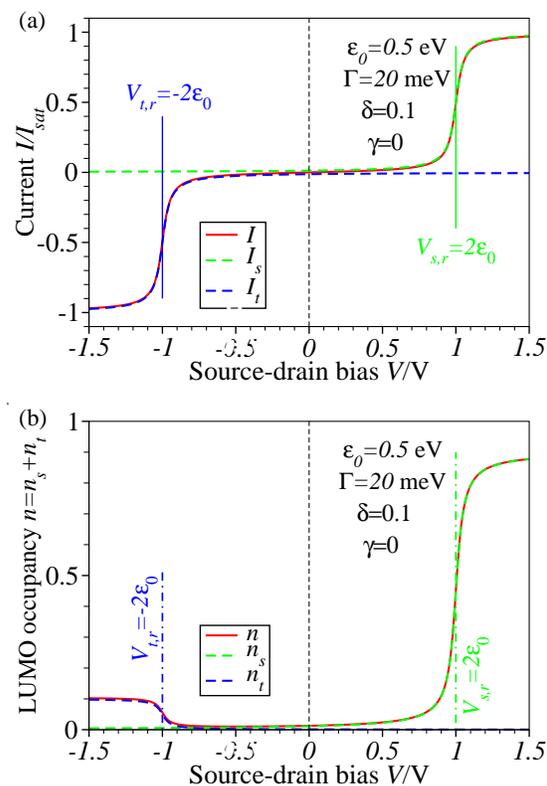

\centerline{\includegraphics[width=0.4\textwidth,angle=0]{fig_iv_delta_0.1_gamma_0.5_new.eps}}
\centerline{\includegraphics[width=0.4\textwidth,angle=0]{fig_ne_nt_delta_0.1_gamma_0.5_new.eps}}
$ $\\[0ex]
\caption{Results for $\varepsilon_0=0.5$\,eV, $\Gamma=20$\,meV, $\delta=0.1$, and 
$\gamma=0$
showing that the biases $V$ where 
steps and plateaus in the current $I$ and LUMO occupancy $n$ occur are controlled
by the values $V_{s,r}$ and $V_{t,r}$ (see eqn~(5) and (6) of the main text) 
where the LUMO 
becomes resonant with the Fermi levels of electrodes $s$(ubstrate) and $t$(ip), 
respectively.
Notice that reduction efficiency (quantified by the LUMO occupancy $n$) is strongly dependent on the 
bias polarity in case of an asymmetric molecule electrode coupling ($\Gamma_s \neq \Gamma_t$),  
unlike the current plateau value, which is unchanged upon bias polarity reversal. 
$n_s$ and $n_t$ represent the separate electrodes' contributions to the LUMO occupancy $n$
expressed by the first and second terms in the RHS of eqn~(2) from the main text.
\ib{For better readability, a vertical dashed line marks the zero bias reference.}
}
\label{fig:delta=0.1_gamma=0.5}
\end{figure}

\begin{figure}[htb]
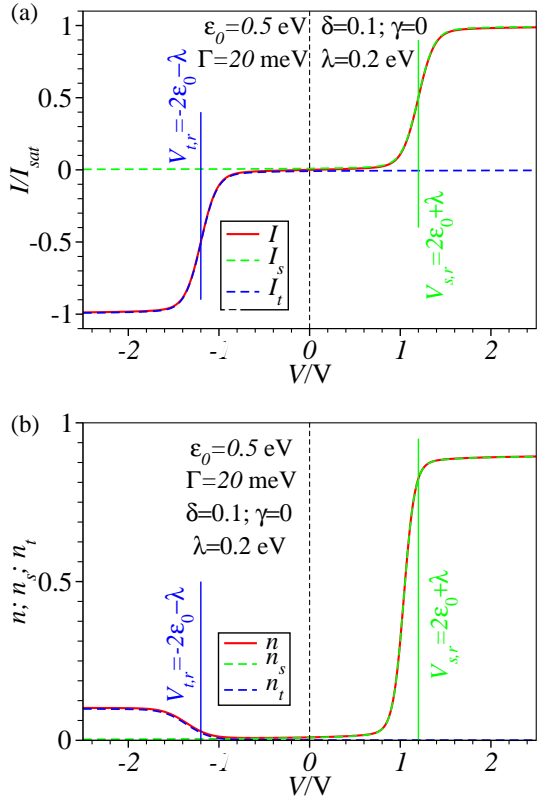

\centerline{\includegraphics[width=0.4\textwidth,angle=0]{fig_iv_delta_0.1_gamma_0.5_lambda_0.2_new.eps}}
\centerline{\includegraphics[width=0.4\textwidth,angle=0]{fig_ne_nt_delta_0.1_gamma_0.5_lambda_0.2_new.eps}}
$ $\\[0ex]
\caption{Results similar to those of \figurename\ref{fig:delta=0.1_gamma=0.5}
including in addition reorganization effects quantified by the reorganization energy
$\lambda$ whose value is given in the legend. 
\ib{For better readability, a vertical dashed line marks the zero bias reference.}
}
\label{fig:delta=0.1_gamma=0.5_lambda_0.2}
\end{figure}

The framework adopted here is provided by the single-level Newns-Anderson model, which was discussed 
\cite{Schmickler:86,Medvedev:07b} and validated for a variety of 
molecular junctions,\cite{Baldea:2012a,Baldea:2012g,Baldea:2013b} 
including those used in the existing SERS-transport studies.\cite{Natelson:14,Baldea:2014f}
We checked that reorganization effects 
\cite{Schmickler:86,Schmickler:97,Kuznetsov:02c,Medvedev:07b,Baldea:2014a,Baldea:2014f} 
do not qualitatively change the present conclusions.
This is illustrated by the comparison between 
\figurename\ref{fig:n_gamma=0.5_0.3_various_delta_vs_V}
and
\figurename\ref{fig:n_gamma=0.5_0.3_various_delta_lambda_0.2_vs_V},
\figurename\ref{fig:delta=0.1_gamma=0.5} and
\figurename\ref{fig:delta=0.1_gamma=0.5_lambda_0.2},
and between panels b and c of \figurename\ref{fig:i_n_vs_e0}.
To make the presentation as simple and clear as possible,
we will give below only formulae wherein reorganization effects are disregarded; 
details on the ensemble averaging needed to include these effects are not given here 
but can be found elsewhere.
\cite{Schmickler:86,Medvedev:07b,Baldea:2014a}

Eqn~(\ref{eq-I}), (\ref{eq-n}), (\ref{eq-gamma}) and (\ref{eq-Gamma-delta}), 
\cite{Schmickler:86,Medvedev:07b,Baldea:2014a} 
which allow to express the current $I$ and the LUMO occupancy $n$, 
constitute the framework of the present discussion 
($e$ and $h$ are the elementary charge and Planck's constant, respectively). 
\begin{eqnarray}
\label{eq-I}
I & = & \frac{I_{sat}}{\pi}  
\left[
\arctan\frac{\varepsilon_{0}(V) + \frac{e V}{2}}{\Gamma} 
- \arctan\frac{\varepsilon_{0}(V) - \frac{e V}{2}}{\Gamma}
\right] \\
\label{eq-I-sat}
I_{sat}  & = &  2 \pi e \Gamma_s \Gamma_t/(h \Gamma \\
\label{eq-gamma}
\varepsilon_{0}(V) & = & \varepsilon_0 + \gamma e V \\
\label{eq-Gamma-delta}
\Gamma_{s} & = & 2\Gamma (1-\delta) \ ; \ \Gamma_{t} = 2 \Gamma \delta 
\end{eqnarray}

The LUMO occupancy $n$ of the molecule embedded in a biased junction 
comprises contributions $n_{s,t}$ from the two ($s$ and $t$) electrodes \cite{Medvedev:07,Baldea:2013d}
\begin{eqnarray}
\label{eq-n}
n & = & n_{s} + n_{t} \\
\label{eq-ns}
n_{s} & = & \frac{1 - \delta}{\pi}
\mbox{arccot}\frac{\varepsilon_{0}(V) - e V/2}{\Gamma} \\
\label{eq-nt}
n_{t} & = & \frac{\delta}{\pi}
\mbox{arccot}\frac{\varepsilon_{0}(V) + e V/2}{\Gamma} 
\end{eqnarray}

The Newns-Anderson model embodies two asymmetries characterized by the dimensionless parameters $\gamma$
and $\delta$ defined in eqn~(\ref{eq-gamma}) and (\ref{eq-Gamma-delta}).
The potential profile asymmetry (or voltage division factor) $\gamma$ quantifies the bias-induced 
shift in LUMO energy offset relative to the Fermi energy ($E_F \equiv 0$) with respect to 
the case of unbiased electrodes $\varepsilon_0 \to \varepsilon_{0}(V)$.\cite{Datta:03,Medvedev:07b,Baldea:2012a}
$\Gamma_{s,t}$ denote the couplings of the molecule to 
generic ``left''/``substrate'' ($s$) and ``right''/``tip'' ($t$) electrodes.
Without loss of generality (because it merely amounts to appropriately label the  
electrodes) we assume that $s$(ubstrate) is not the weakest coupled electrode to the LUMO
($\Gamma_t \leq \Gamma_s$, $0 < \delta \leq 1/2$)
and define the source-drain bias $V$ with respect to its potential ($V=\mathcal{V}_{t} - \mathcal{V}_{s}$).
Using a potential origin located symmetrically between the electrodes' Fermi energies 
($-e \mathcal{V}_{s,t} = \pm e V/2$), 
$-1/2 < \gamma < 1/2$. $\gamma$ is positive (negative) if the bias 
shifts the LUMO towards the Fermi level of electrode $s$ ($t$). 

Similar to $I$ 
(\figurename\ref{fig:delta=0.1_gamma=0.5} and \figurename\ref{fig:delta=0.1_gamma=0.5_lambda_0.2}),\cite{Baldea:2014f} 
the LUMO occupancy exhibits substantial changes within 
ranges $\delta V \sim \Gamma$ (which are narrow in typical non-resonant
cases $\Gamma \ll \varepsilon_0$) at the biases (case $\lambda=0$)
\begin{eqnarray}
\label{eq-V_rs}
V_{r,s} & = & \varepsilon_{0}/(1/2 - \gamma) \\
\label{eq-V_rt}
V_{r,t} & = & - \varepsilon_{0}/(1/2 + \gamma) 
\end{eqnarray}
where the LUMO and electrodes' Fermi level become resonant and plateaus (saturation) above these biases.
The saturation values (subscript $sat$) can be expressed as follows (superscripts $\pm$ refer to bias polarity)
\begin{eqnarray}
\label{eq-ns-max}
n 
& \xlongrightarrow[\text{large positive $V$ }]{\text{$V - V_{r,s}\gg \Gamma$}} & 
n_{sat}^{+} 
\approx 1 - \delta  \\
\label{eq-nt-max}
n 
& \xlongrightarrow[\text{large negative $V$ }]{\text{$\vert V - V_{r,t} \vert \gg \Gamma$}} & 
n_{sat}^{-}  
\approx \delta  
\end{eqnarray}
\section{Results and discussion}
\label{sec:results}
\begin{figure}[htb]
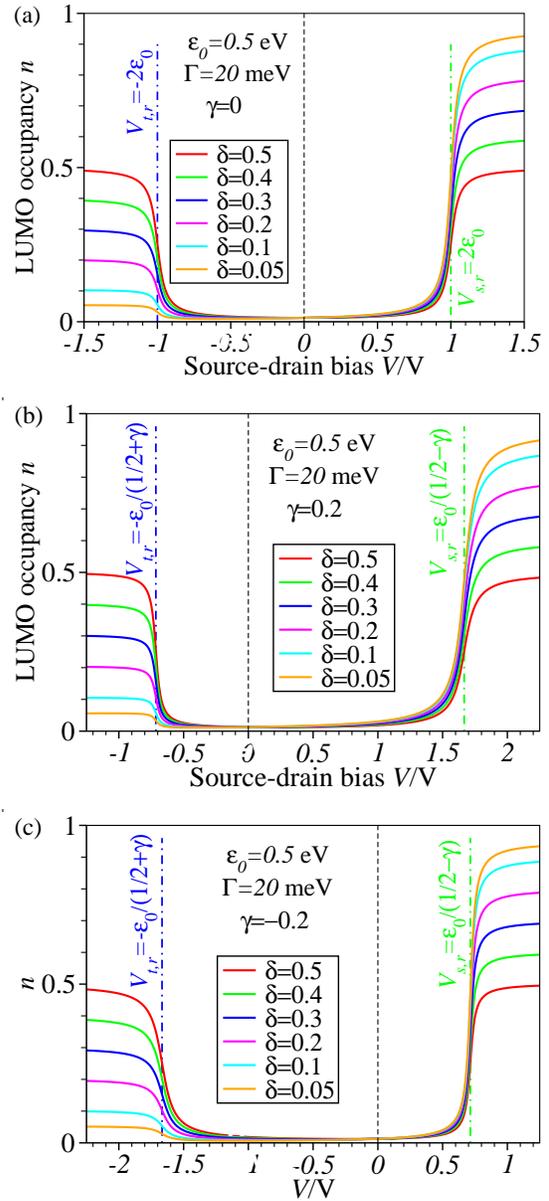

\centerline{\includegraphics[width=0.4\textwidth,angle=0]{fig_n_various_delta_gamma_0.5_vs_V_new.eps}}
\centerline{\includegraphics[width=0.4\textwidth,angle=0]{fig_n_various_delta_gamma_0.3_vs_V_new.eps}}
\centerline{\includegraphics[width=0.4\textwidth,angle=0]{fig_n_various_delta_gamma_0.7_vs_V_new.eps}}
$ $\\[0ex]
\caption{The stronger the asymmetry of molecule-electrode couplings (smaller $\delta$'s), 
the more efficient is the LUMO reduction. 
Notice that a nonvanishing value of $\gamma$
(panels b and c) breaks the bias reversal symmetry ($n(-V) \neq n(V)$) but does not change 
the plateau values of $n$ with respect to the case $\gamma=0$ (panel a).
The values of $\varepsilon_0$ and $\Gamma$ are comparable to those used in existing 
concurrent SERS-transport studies.\cite{Natelson:14,Baldea:2014f} 
\ib{For better readability, a vertical dashed line marks the zero bias reference.}
}
\label{fig:n_gamma=0.5_0.3_various_delta_vs_V}
\end{figure}

\begin{figure}[htb]
\centerline{\includegraphics[width=0.4\textwidth,angle=0]{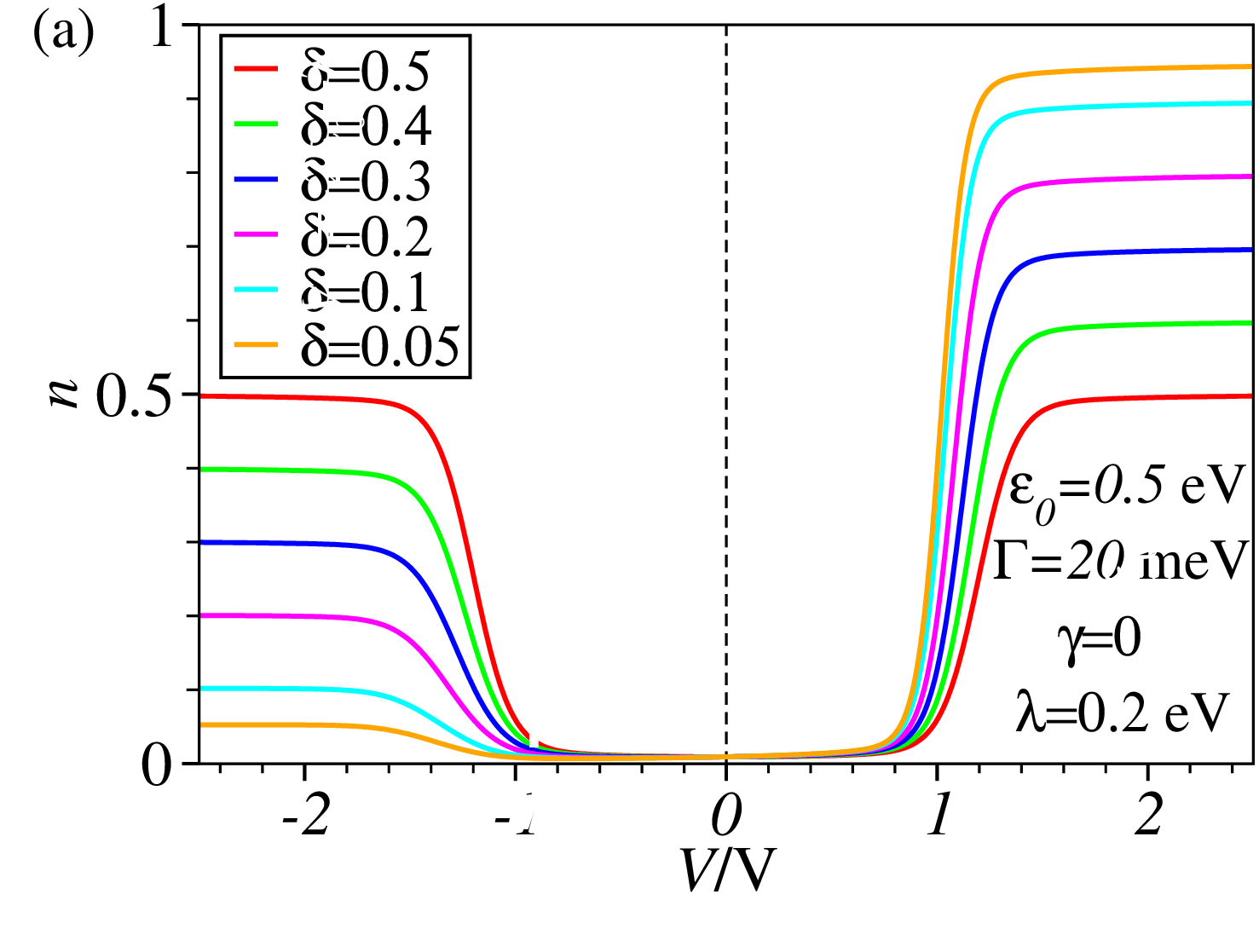}}
\centerline{\includegraphics[width=0.4\textwidth,angle=0]{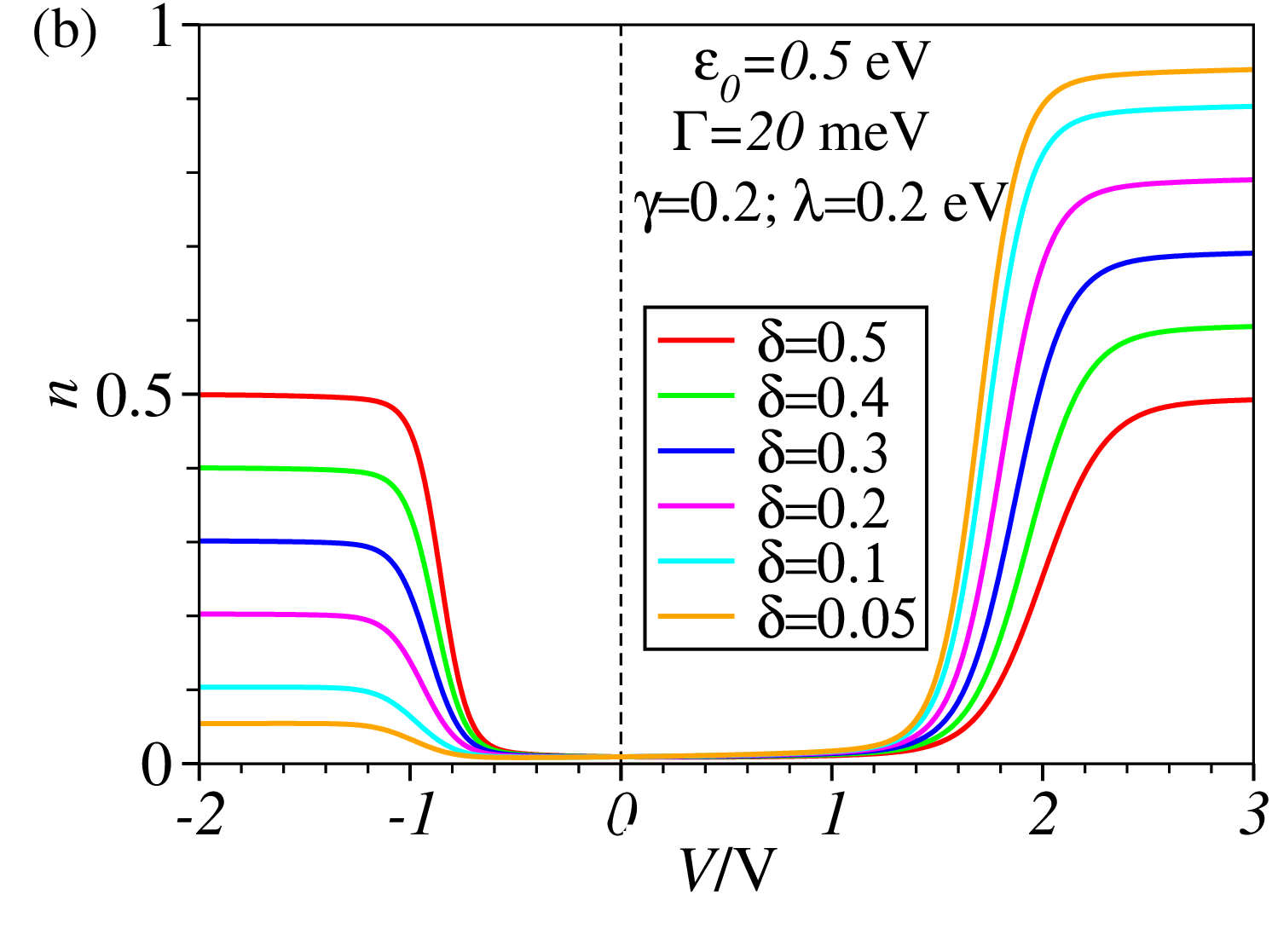}}
\centerline{\includegraphics[width=0.4\textwidth,angle=0]{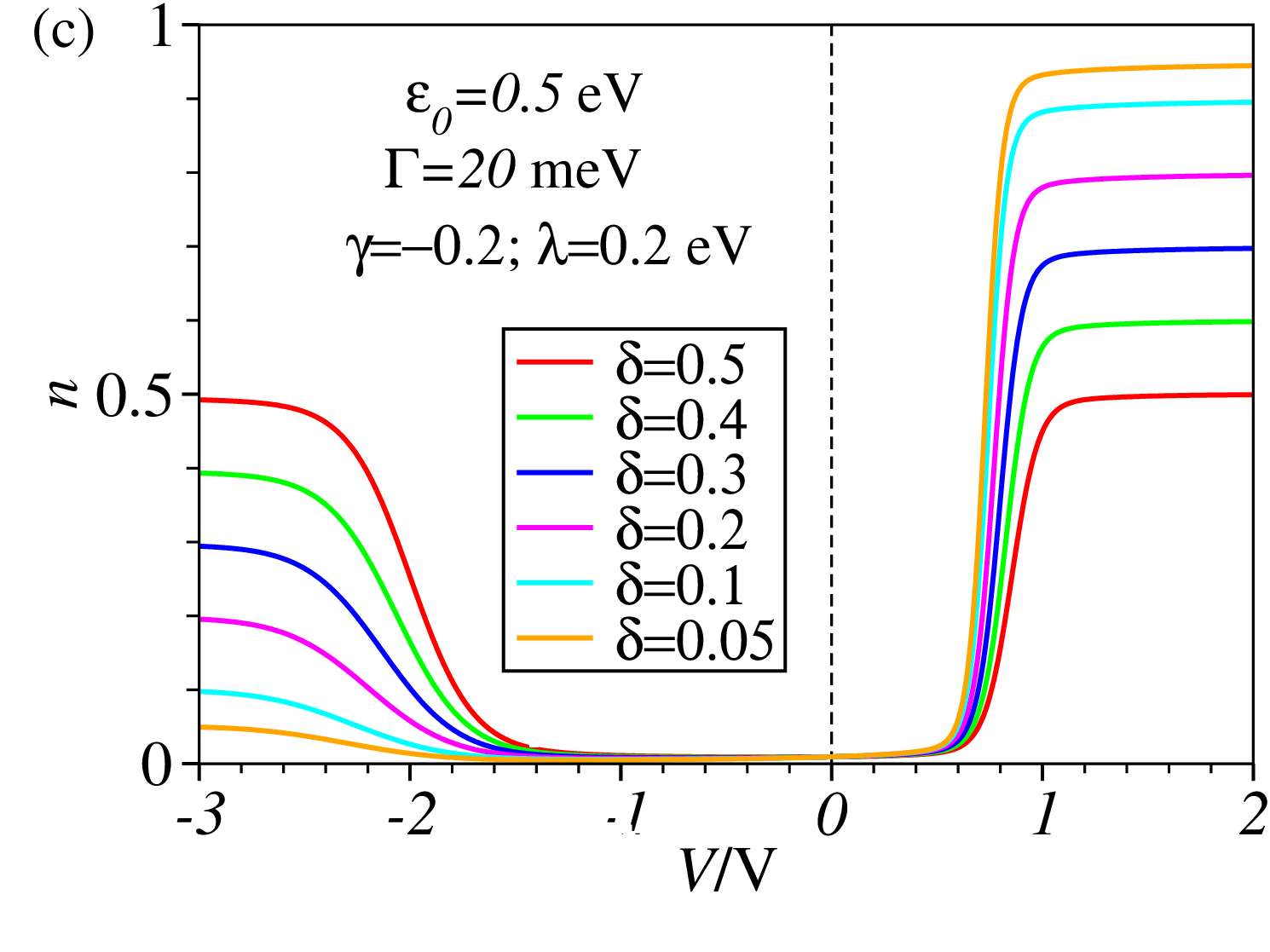}}
$ $\\[0ex]
\caption{Results similar to those of \figurename\ref{fig:n_gamma=0.5_0.3_various_delta_vs_V}
including in addition reorganization effects quantified by the reorganization energy
$\lambda$ whose value is given in the legend. 
\ib{For better readability, a vertical dashed line marks the zero bias reference.}
}
\label{fig:n_gamma=0.5_0.3_various_delta_lambda_0.2_vs_V}
\end{figure}

The impact of the source-drain bias on the molecular charge is depicted in 
\figurename\ref{fig:n_gamma=0.5_0.3_various_delta_vs_V} and 
\figurename\ref{fig:n_gamma=0.5_0.3_various_delta_lambda_0.2_vs_V}.
It is worth noting that 
the asymmetry in the electronic coupling ($\delta \neq 1/2$, $\Gamma_s \neq \Gamma_t$) 
is physically distinct
from the asymmetry ($\gamma \neq 0$) in the electric potential drop across a molecular junction.
\cite{Baldea:2012a,Baldea:2014f,Baldea:2014g}  
Their impact on $n$ is completely different. The comparison between 
\figurename\ref{fig:n_gamma=0.5_0.3_various_delta_vs_V}a and b
reveals that $\gamma \neq 0$ merely 
breaks the bias reversal symmetry ($n(V) \neq n(-V)$). This is the counterpart of the current rectification 
($I(-V) \neq -I(V)$, \emphj{cf.}~eqn~(\ref{eq-I}) and (\ref{eq-gamma})).
As long as the molecule-electrode coupling is symmetric ($\delta = 1/2$), the plateau values at either 
bias polarity are equal, $n_{sat}^{+} = n_{sat}^{-} = 1/2$.

From the perspective of achieving a full redox process in biased molecular junctions, 
this result for symmetric coupling is disappointing: whatever high is the source-drain 
bias which a molecular junction
can withstand ($\vert V\vert \gg V_{r,s} \mbox{ or } \vert V_{r,t}\vert$), 
whether the potential profile is symmetric ($\gamma=0$) or asymmetric ($\gamma \neq 0$),
the average molecular charge cannot exceed half of an electron 
($n_{sat}^{\pm} = 1/2$ for $\delta=1/2$, \emphj{cf.}~eqn~(\ref{eq-n})). 
This is a surprising result; intuitively,
one may expect a full change in the molecular redox state ($n=1$) at biases
corresponding to resonant transport 
($\varepsilon_{0}(V) \approx \mu_{s}(V) \mbox{ or } \mu_{t}(V)$).\cite{Natelson:14}
In fact, exactly on resonance 
($V=V_{r,s} \mbox{ or } V_{r,t}$) the average molecular charge is even smaller: 
$n\approx 1/4$, see eqn~(\ref{eq-ns-r}) and (\ref{eq-ns-r}) for $\delta=1/2$, $\Gamma \ll \varepsilon_0$).
\begin{eqnarray}
\label{eq-ns-r}
n(V=V_{r,s}) & = & 
\underbrace{\frac{1 - \delta}{2}}_{n_s} + \underbrace{\frac{\delta}{\pi} \arctan\frac{\Gamma}{eV_{r,s}}}_{n_t} 
\raisebox{1.8ex}{ $\underrightarrow{\Gamma \ll \varepsilon_0}$ }
\frac{1 - \delta}{2} 
\\
\label{eq-nt-r}
n(V=V_{r,t}) & = & 
\underbrace{\frac{1 - \delta}{\pi} \arctan\frac{\Gamma}{e\vert V_{r,t}\vert}}_{n_s} + \underbrace{\frac{\delta}{2}}_{n_t} 
\raisebox{1.8ex}{ $\underrightarrow{\Gamma \ll \varepsilon_0}$ }
\frac{\delta}{2} 
\end{eqnarray}

\begin{figure}[htb]
$ $\\[-8ex]
\centerline{\includegraphics[width=0.35\textwidth,angle=0]{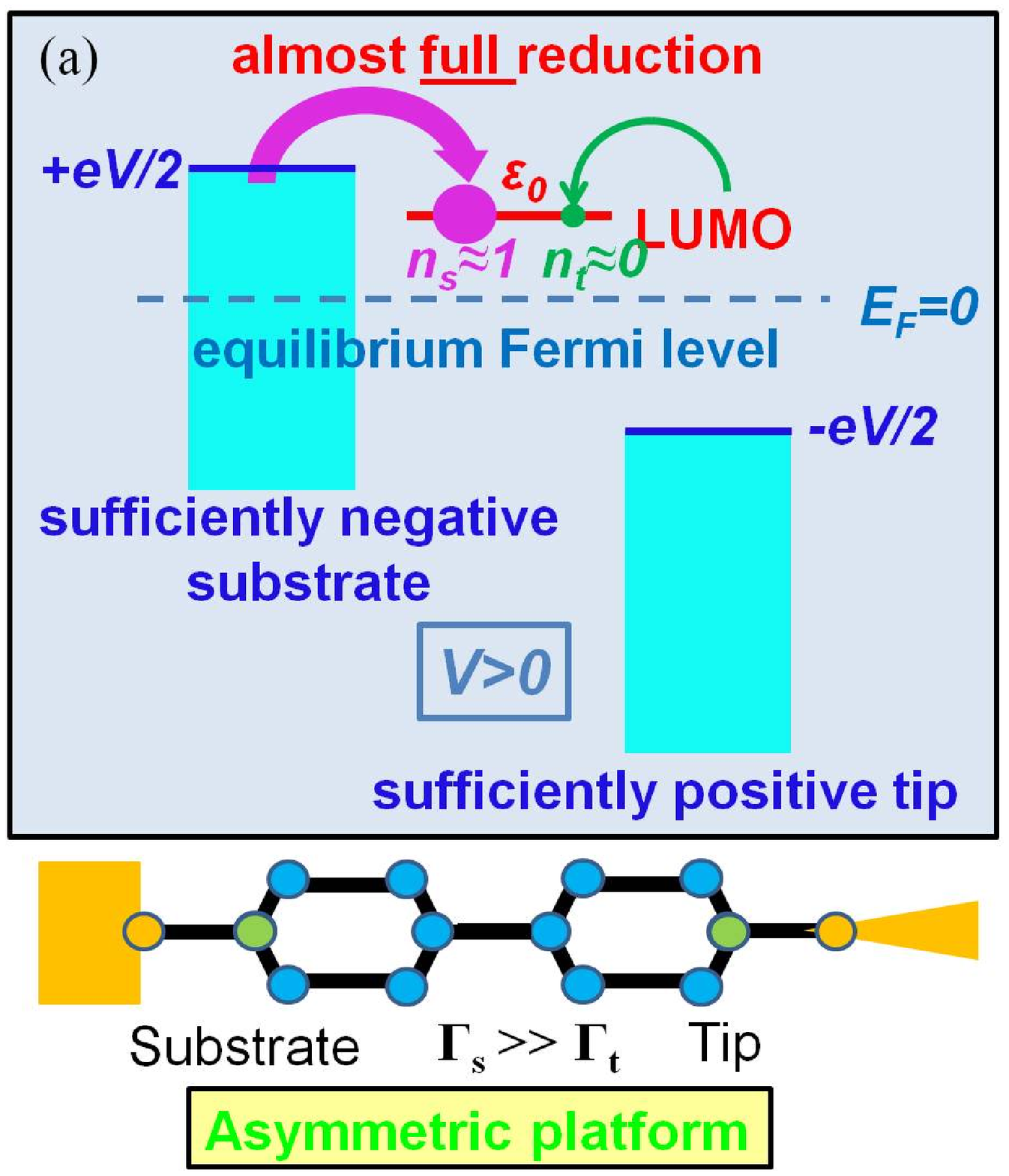}}
$ $\\[-16ex]
\centerline{\includegraphics[width=0.35\textwidth,angle=0]{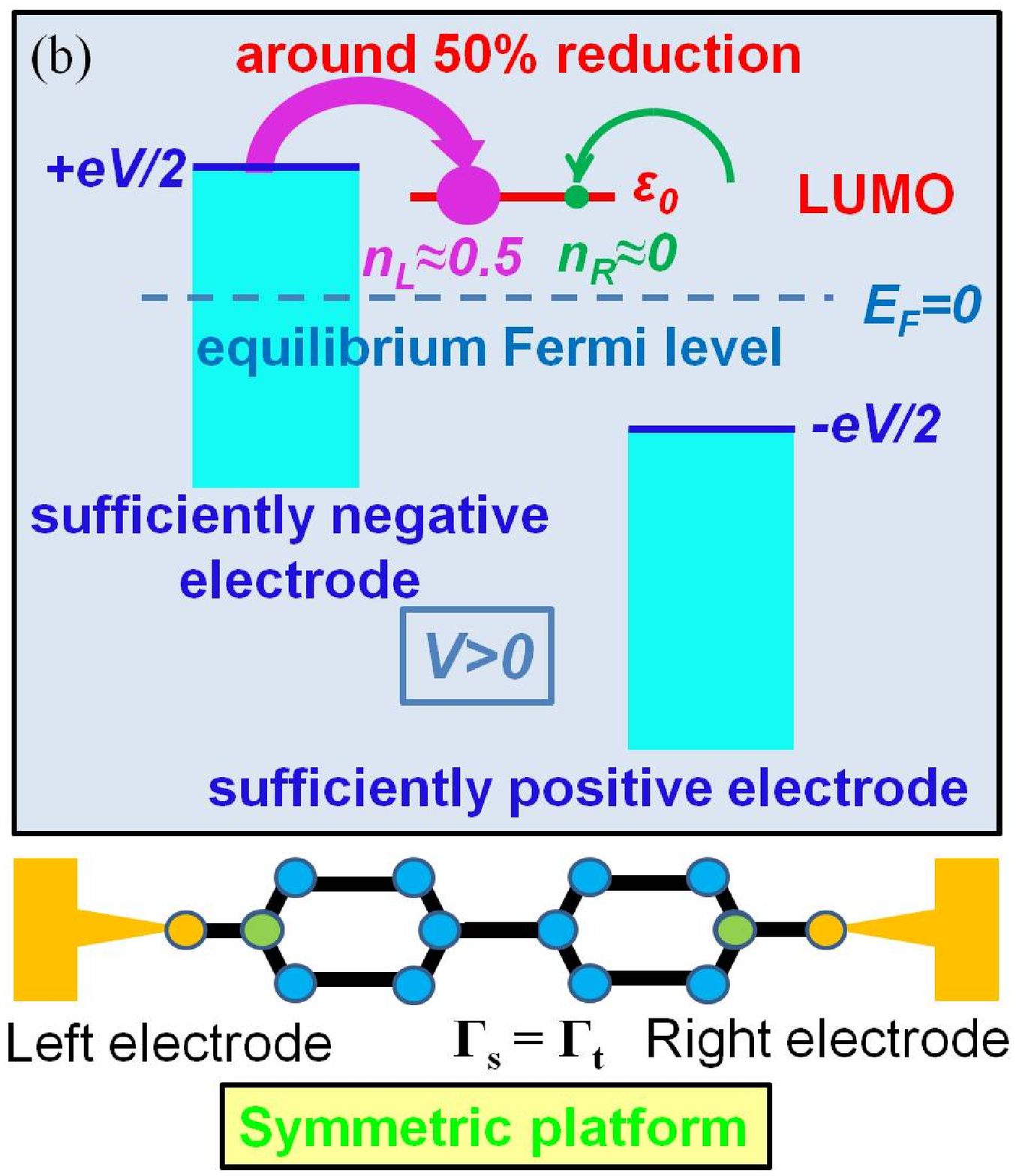}}
$ $\\[-16ex]
\caption{An almost complete reduction can be achieved if the LUMO is much stronger coupled 
to one electrode ($\Gamma_s \gg \Gamma_t$) having a sufficiently negative potential (panel a),
which is impossible for symmetric coupling to electrodes (panel b).
}
\label{fig:ReductionAsymmetricPlatform}
\end{figure}

According to eqn~(\ref{eq-ns-max}) and (\ref{eq-nt-max}), 
a molecule in a current-carrying state can accommodate an average charge larger that $n=1/2$
for junctions characterized by asymmetric molecule-electrode couplings $\delta \neq 1/2$.
The stronger the asymmetry, the larger is the charge of the average redox state 
(\figurename\ref{fig:n_gamma=0.5_0.3_various_delta_vs_V}).
An almost perfect reduced state ($n\approx 1$) can be achieved  
for highly asymmetric
molecule-electrode couplings ($\delta \ll 1$) 
if the potential of the strongest coupled electrode ($s$) is sufficiently negative
(sufficiently large positive biases). 
In such cases ($\Gamma_s \gg \Gamma_t$ and $V - V_{r,s} \gg \Gamma/e$), 
the Fermi energy of electrode $s$ is sufficiently above the LUMO energy, and the electron
transferred from this electrode with a high 
rate ($\Gamma_s$) almost entirely remains on the LUMO in cases where the 
transfer from the LUMO to electrode $t$ is inefficient (small $\Gamma_t$):
$n \approx \Gamma_s/\left(\Gamma_s + \Gamma_t\right) = 1 - \delta \alt 1$ 
(\figurename\ref{fig:ReductionAsymmetricPlatform}a). This is impossible in cases 
of symmetric coupling ($\Gamma_s = \Gamma_t$), 
wherein $n \approx \Gamma_s/\left(\Gamma_s + \Gamma_t\right) = 1/2$; 
on average, half of the entire electron transferred from electrode $s$ to 
the LUMO is further transferred to electrode $t$ in cases wherein the two transfer rates are 
equal (\figurename\ref{fig:ReductionAsymmetricPlatform}b).

The foregoing analysis demonstrated that an almost complete reduction 
can be obtained (i) for a molecule highly asymmetrically coupled to 
electrodes and (ii) sufficiently high biases. 
While the first condition ($\Gamma_t \ll \Gamma_s$) 
can be easily satisfied in asymmetrical
\ib{EC}-STM (see Sec.~\ref{sec:em-vs-stm}) \cite{Baldea:2013d,Baldea:2014f} 
setups, the second condition seems problematic. 
\figurename\ref{fig:delta=0.1_gamma=0.5} 
illustrates that the biases needed for an almost perfect reduction  
are those where current plateaus occur.
 
As experiments on molecular junctions
did not routinely report currents exhibiting plateaus with increasing bias
(we are aware of one exception \cite{Metzger:01}), this appears to be 
an important practical limitation.

\begin{figure}[htb]
$ $\\[-16ex]
\centerline{\includegraphics[width=0.35\textwidth,angle=0]{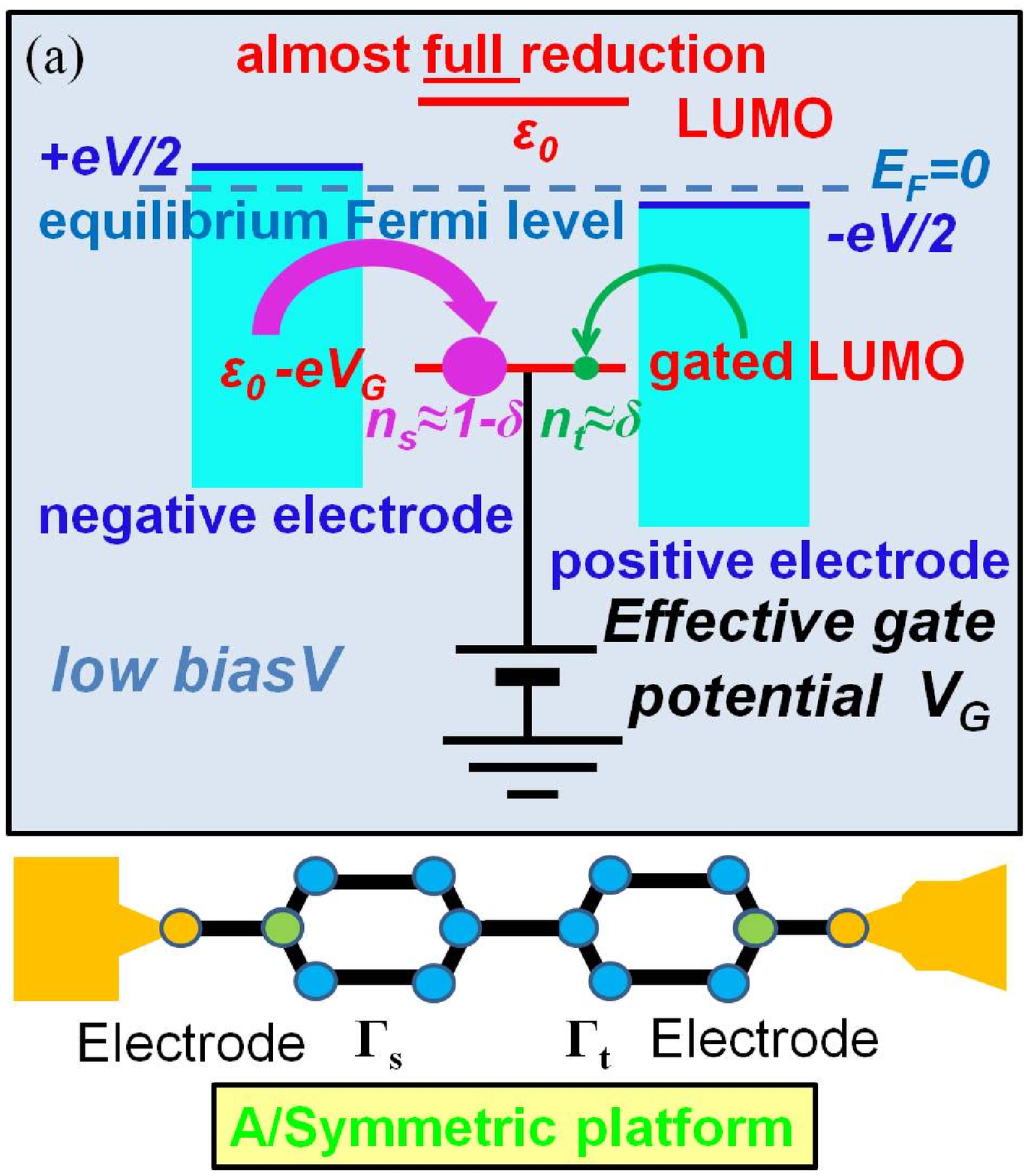}}
$ $\\[-5ex]
\centerline{\includegraphics[width=0.35\textwidth,angle=0]{fig_n_various_delta_gamma_0.5_vs_e0.eps}}
\centerline{\includegraphics[width=0.35\textwidth,angle=0]{fig_ie0_delta_0.5_gamma_0.5_V_0.05_lambda_0.2_vs_e0_pccp.eps}}
$ $\\[-2ex]
\caption{Results demonstrating that an efficient reduction 
can be achieved by lowering the LUMO energy below the electrodes' Fermi level
via LUMO gating (panel a). An almost complete reduction (values of $n$ close to unity)
can be obtained for reasonable gate-driven LUMO shifts;
in this case, the coupling asymmetry $\delta$ plays a secondary role:
$n \approx 95$\% (almost $\delta$-independent) at $\varepsilon_0 \approx -0.1$\,eV 
(without reorganization effects: panel b; with reorganization effects: panel c).
\ib{These results emphasize that it is not the (source-drain) bias $V$, 
but rather the overpotential (on which $\varepsilon_0$ linearly depends) 
that determines the LUMO occupancy $n$ in electrochemical environment.} 
}
\label{fig:i_n_vs_e0}
\end{figure}

A nearly complete reduction would be possible if the LUMO 
lied below the Fermi level of the strongest coupled electrode
($n \approx 1 - \delta$ for $\delta \ll 1$ form eqn~(\ref{eq-n}))
or below the Fermi levels of both electrodes (\figurename\ref{fig:i_n_vs_e0}). 
This results from eqn~(\ref{eq-n}): 
$n \approx (1 - \delta) + \delta = 1$ for $\varepsilon_0 < 0$
(LUMO below the Fermi level $E_F$($=0$) of unbiased electrodes).
In the absence of any bias, $\varepsilon_0 > 0$; 
the Fermi level lies within the HOMO-LUMO gap 
of a molecule linked to electrodes (charge neutrality).
However, $\varepsilon_0 < 0$ becomes possible under electrostatic gating.
\cite{Tao:96,Alessandrini:06,Reed:09,Wandlowski:08}
An appropriate gate potential $V_G$ 
\ib{(overpotential in electrochemical language, on which the LUMO energy $\varepsilon_0$ linearly 
depends\cite{Alessandrini:06,Medvedev:07,Wandlowski:08,Baldea:2013d})}
can lower the LUMO energy below $E_F$ (\figurename\ref{fig:i_n_vs_e0}a).
$I-V_G$ ($I-\varepsilon_0$) transfer characteristics exhibiting maxima, which occur at  
resonance ($\varepsilon_0 \approx 0$),\cite{Medvedev:07b,Wandlowski:08,Baldea:2014a} can be taken as indicating
a substantial change in the molecular redox state 
(from $n \approx 0 \mbox{ for } \varepsilon_0 > 0 \mbox{ to } n \approx 1 \mbox{ for } \varepsilon_0 < 0$).
\ib{This is illustrated in \figurename\ref{fig:i_n_vs_e0}b, 
which emphasizes that it is not the (source-drain) bias $V$, 
but rather the overpotential that determines the reduction efficiency in electrochemical environment.} 
\ib{
\section{Specific remarks}
\label{sec:remarks}
In this section we will address three specific issues related to the results reported above.
\subsection{Rectification \emph{vs.}~reduction}
\label{sec:I-vs-na}
To avoid possible misunderstandings, in this subsection we briefly 
discuss some of 
the present results in the context of those reported previously. 

In agreement with previous work 
(\emph{e.g.}, ref.~\citenum{Baldea:2012b} and \citenum{Ratner:15a}),
eqn (\ref{eq-I}), (\ref{eq-I-sat}) and (\ref{eq-gamma}) show that 
asymmetric molecule-electrode couplings \emph{alone} (\emph{i.e.}, 
$\Gamma_s \neq \Gamma_t$ (or $\delta \neq 1/2$) \emph{and} $\gamma=0$)
do \emph{not} yield rectification: the aforementioned equations 
yield $I(-V) = - I(V)$ for $\gamma = 0$ irrespective of the value of $\delta$. 

The fact that a potential profile asymmetry ($\gamma \neq 0$) 
yields current rectification has been amply discussed. A very incomplete list includes
ref.~\citenum{Metzger:01b,Metzger:01,Baldea:2012a,Ratner:15a}; notice that the parameters 
$p$ of ref.~\citenum{Metzger:01b,Metzger:01} and $a$ of ref.~\citenum{Ratner:15a}
correspond to that denoted by $\gamma$ in this paper. In particular, 
the present eqn~(\ref{eq-I}) coincides with eqn (4b) of ref.~\citenum{Ratner:15a}.
This is why all conclusions on the current asymmetry (``current rectification'') emerging from our 
eqn~(\ref{eq-I}) are not new; they coincides with those of the earlier works 
based on the same formula for the current $I=I(V)$.  

Important new results reported in the present paper are that an applied bias 
can yield an asymmetric molecular reduction $n(-V) \neq n(V)$ 
and that current asymmetry $I(-V) \neq - I(V)$ and reduction asymmetry 
$n(-V) \neq n(V)$ are conceptually different. In particular, a
symmetric current $I(-V) = -I(V)$ (no rectification) does not rule out 
an asymmetric LUMO occupancy $n(-V) \neq n(V)$ 
(\emphj{cf.}~\figurename\ref{fig:delta=0.1_gamma=0.5})
and current rectification $I(-V) \neq -I(V)$ does not rule out symmetric 
plateau values of the reduction degree 
($n(V\ll - 2\varepsilon_0) = n(V \gg 2\varepsilon_0) = 1/2$ for $\Gamma_s = \Gamma_t$).

Concerning the current rectification a final comment is
in order, however. Similar to ref.~\citenum{Baldea:2012b} and \citenum{Ratner:15a},
eqn (\ref{eq-I}), (\ref{eq-I-sat}) and (\ref{eq-gamma}) refer to situations wherein
reorganization effects are ignored (\emph{i.e.}, $\lambda \equiv 0$). 
As recent work demonstrated,\cite{Baldea:2013d} 
in cases where reorganization effects are non-negligible ($\lambda \neq 0$),
the asymmetry $\Gamma_s \neq \Gamma_t$
yields current rectification ($I(-V) \neq -I(-V)$) \emph{even} 
if there is no bias-induced 
shift of the LUMO energy ($\gamma = 0$); see eqn (13), and (18) to (23) of 
ref.~\citenum{Baldea:2013d}.\cite{rectification-invisible} 
Without intending to 
be exhaustive (current rectification in not our primary aim here),
we mention that,
in cases wherein $\Gamma_s \neq \Gamma_t$ and $\gamma = 0$,
current rectification may also appear due to charging effects.\cite{Datta:04d,relation-to-datta}   
\subsection{Symmetric \emphj{vs.}~asymmetric contact couplings and experimental platforms}
\label{sec:em-vs-stm}
The above analysis demonstrated the key role played by the molecule-electrode couplings
($\Gamma_{x}$, $x=s,t$) 
in determining the reduction efficiency. For electrodes with wide flat band structures 
around the Fermi energy, $\Gamma_{s,t}$ 
can be expressed in terms of the electrode density of states at the
Fermi level $\rho_{s,t}$ and effective transfer integrals $t_{s,t}$ quantifying the charge transfer
between electrodes and the dominant molecular orbital (LUMO in the specific case considered here)
\cite{CuevasScheer:10}
\begin{equation}
\label{eq-wbl}
\Gamma_{x} = \rho_{x} t_{x}^{2}
\end{equation}

Let us now consider the experiment that succeeded to reveal a change in the redox state of a 
\ce{C60} molecule embedded in a biased junction by means of simultaneous SERS-transport measurements in 
an electromigration platform.\cite{Natelson:14} In view of the high molecular symmetry and 
of the (nearly) symmetric electrodes characterizing electromigrated junctions, one can assume
$\rho_{s} \approx \rho_{t}$ and $t_{s} \approx t_{t}$. This implies $\Gamma_{s} \approx \Gamma_{t}$,
and we have shown above that in this symmetric case ($\delta=1/2$) reduction cannot exceed 50\%.
Drawing (experimentalists') attention on the limited reduction degree that can be achieved even if 
such electromigrated molecular junctions can be brought into a current plateaus regime represents an important
aim of the present theoretical work.\cite{ignore-threefold-lumo-degeneracy} 

In typical STM or CP-AFM experiments, molecules forming regular self-assembled monolayers 
(SAMs) are (covalently) bound to the substrate.
A difference $\rho_s \neq \rho_t$ (yielding $\Gamma_s \neq \Gamma_t$ via 
eqn~(\ref{eq-wbl})) may exist because the substrate surface is 
typically monocrystalline (\emphj{e.g.}, Au(111)) while the tip facet is undefined.
Still, it is more probable that not the difference $\rho_s \neq \rho_t$ but rather 
that of the charge transfer efficiency 
($t_s \neq t_t$) determines the contact coupling asymmetry $\Gamma_s \neq \Gamma_t$.

Quantifying the asymmetry $\Gamma_s \neq \Gamma_t$ from transport measurements is not straightforward;
we have seen above that
(nearly) symmetric curves $I(-V) \approx -I(V)$, which are measured in numerous junctions based on symmetric 
(and occasionally also asymmetric) molecules,  
do not exclude (highly) asymmetric contact couplings. 

A coupling asymmetry $\Gamma_s > \Gamma_t$ has been concluded in ref.~\citenum{Vuillaume:15b} 
after a detailed analysis of the transport data in CP-AFM junctions. 
In CP-AFM setups such an asymmetry appears to be plausible because one can expect that $t_s > t_t$, 
given the fact that a stable molecule-tip chemical bond is hard to imagine; normally, 
charge transport only occurs by applying a loading force at the CP-AFM tip.\cite{Frisbie:00}
A similar inequality ($t_s > t_t$, $\Gamma_s > \Gamma_t$) can also be expected for STM break junctions;
the formation of a stable molecule-tip chemical bond is implausible 
during repeated processes of rapidly crashing the tip into and retracting it away 
from the substrate.\cite{Tao:03,Guo:11} So, although \emphj{per se} the asymmetric outlook of the 
STM and CP-AFM setups does not necessarily imply $\Gamma_s > \Gamma_t$, this inequality can be expected
in view of the different bond strength at the contacts. 
The fact that the symmetry of the electromigration platform and the asymmetry 
of the STM platform do not merely refer to the usual 
schematic illustrations of these setups (like those in \figurename\ref{fig:ReductionAsymmetricPlatform}
and \figurename\ref{fig:i_n_vs_e0}), but also have a physical content has been 
recently quantified.\cite{Baldea:2014e,symmetric-molecules}

Out of the various experimental platforms employed, EC-STM setups appears to be the most favorable, 
enabling almost perfect reduction.
In electrochemical environment (EC-STM),\cite{Wandlowski:08}
the tip typically approaches but does not contact the free end of the molecules.  
The different through-bond \emphj{vs.}~through-space charge transfer mechanisms 
at the (EC-)STM substrate and tip
reflect themselves in significantly different transfer integrals ($t_s \gg t_t$)
responsible for the high asymmetry $\Gamma_s \gg \Gamma_t$ (\emphj{cf.}~eqn~(\ref{eq-wbl}));
in agreement with this analysis,
our recent works demonstrated very highly asymmetric molecule-electrode couplings:
$\delta \sim 10^{-4}$ (ref.~\citenum{Baldea:2013d}) and 
$\delta \sim 10^{-2}$ (ref.~\citenum{Baldea:2014f}). 
\subsection{The Newns-Anderson model}
\label{sec:na}
As noted in Introduction, the present study has been motivated by the findings of 
ref.~\citenum{Natelson:14} and \citenum{Baldea:2014f}, which demonstrated that reduction  
is possible in molecular junctions in current carrying states 
and concluded that the Newns-Anderson model (single level + Lorentzian transmission)
represents an adequate theoretical framework. In favor of the 
the Newns-Anderson model one can still add its
ability to excellently describe the charge transport by tunneling in a variety of 
molecular junctions.\cite{Baldea:2012a,Baldea:2012b,Baldea:2012g,Baldea:2013b} 
The aforementioned represent a sufficient justification for adopting 
the Newns-Anderson model in the present study.

Still, for reasons delineated below we believe that our main conclusion on the 
impact of coupling asymmetry ($\Gamma_s \neq \Gamma_t$) on the reduction efficiency 
holds beyond the Newns-Anderson framework.

Using the expressions of the partial LUMO occupancies $n_s$ and $n_t$ given in eqn~(\ref{eq-ns}) and (\ref{eq-nt}),
respectively, eqn~(\ref{eq-I}) can be recast as
\begin{eqnarray}
\label{eq-I-n}
I & = & e \left(\frac{n_s}{\tau_t} - \frac{n_t}{\tau_s} \right) \\
\label{eq-tau-s,t}
\tau_{s,t} & = & \frac{\hbar}{2\Gamma_{s,t}}
\end{eqnarray}
In eqn~(\ref{eq-I-n}) (where the factor 2 is the spin contribution), 
the first term refers to an electron 
arriving at the LUMO as a result of the coupling of the molecule 
to the left electrode. Eqn~(\ref{eq-ns}) and (\ref{eq-Gamma-delta})
yield
\begin{equation}
\label{eq-ns-Gamma}
n_{s} = \frac{\Gamma_s}{2\Gamma}\frac{1}{\pi}\mbox{arccot}\frac{\varepsilon_{0}(V) - e V/2}{\Gamma}
\propto \Gamma_{s} 
\end{equation}
This electron is transferred to the right electrode
within a characteristic time $\tau_{t}$. Via eqn~(\ref{eq-tau-s,t}), $\tau_{t}$ corresponds to the
rate $\Gamma_{t}$ determined by the coupling to the right electrode. 
Likewise, the second term of eqn~(\ref{eq-I-n}) describes the electron flow in opposite direction:
an electron
arriving at the LUMO as a result of the coupling of the molecule
to the right electrode. Eqn~(\ref{eq-nt}) and (\ref{eq-Gamma-delta}) yield
\begin{equation}
\label{eq-nt-Gamma}
n_{t} = \frac{\Gamma_t}{2\Gamma}\frac{1}{\pi}\mbox{arccot}\frac{\varepsilon_{0}(V) + e V/2}{\Gamma}
\propto \Gamma_{t} 
\end{equation}
This electron is transferred to the left electrode
within a characteristic time $\tau_{s}$. Via eqn~(\ref{eq-tau-s,t}), $\tau_{s}$ corresponds to the
rate $\Gamma_{s}$ determined by the coupling to the left electrode.

Eqn~(\ref{eq-ns-Gamma}) and (\ref{eq-nt-Gamma}) yield the following limiting (plateau) values
\begin{eqnarray}
\label{eq-ns-Gamma-sat}
n 
& \xlongrightarrow[\text{large positive $V$ }]{\text{$V - V_{r,s}\gg \Gamma$}} & 
n_{sat}^{+} \approx n_{s}^{+} 
\approx \frac{\Gamma_s}{2\Gamma}  
\approx \left\{
\begin{array}{l}
\xlongrightarrow[\text{ }]{\text{$\Gamma_s \gg \Gamma_t$}} 1 \\
\xlongrightarrow[\text{ }]{\text{$\Gamma_s \approx \Gamma_t$}} \frac{1}{2} \\
\end{array}
\right . 
\\
\label{eq-nt-Gamma-sat}
n 
& \xlongrightarrow[\text{large negative $V$ }]{\text{$\vert V - V_{r,t} \vert \gg \Gamma$}} & 
n_{sat}^{-}  \approx n_{t}^{+} 
\approx \frac{\Gamma_t}{2\Gamma}  
\approx \left\{
\begin{array}{l}
\xlongrightarrow[\text{ }]{\text{$\Gamma_s \gg \Gamma_t$}} 0 \\
\xlongrightarrow[\text{ }]{\text{$\Gamma_s \approx \Gamma_t$}} \frac{1}{2} \\
\end{array}
\right . 
\end{eqnarray}
It is noteworthy that both the current, eqn~(\ref{eq-I-n}), and the plateau LUMO occupancies,
eqn~(\ref{eq-ns-Gamma}) and (\ref{eq-nt-Gamma}) can be entirely expressed in terms of the rates
$\Gamma_{s,t}$. Parenthetically, these equations also holds in the presence of reorganization ($\lambda \neq 0$).
In view of the appealing simplicity of the above expressions and their clear physical 
content, we believe that they  
apply beyond the Newns-Anderson framework, as also suggested by the cartoons presented in 
\figurename\ref{fig:ReductionAsymmetricPlatform} and \ref{fig:i_n_vs_e0}. 

Obviously, the aforementioned should not be taking as attempting to discourage
alternative approaches of redox processes in molecular junctions based on other 
theoretical models utilized in the literature \cite{CuevasScheer:10}
or by further refinements of the Newns-Anderson model itself. 
Concerning the latter possibility, an extension that appears to us as particularly important 
in studying charge transport through redox units 
is to consider population dependent contact couplings $\Gamma_{s,t} = \Gamma_{s,t}(n)$ 
and LUMO energy $\varepsilon_0 = \varepsilon_{0}(n)$. 

From numerous studies on redox electrochemical systems,\cite{Bard:80} 
it is known that changes in the reduction (or oxidation) degree --- with accompanying
redistributions of the electronic charge over the whole molecule --- have an overall, although 
\emph{selective}
impact on the chemical bond strengths between the various molecular constituents and molecular orbital energies. 
For molecular junctions, 
the influence of the reduction (or oxidation) degree on the bond strengths at the contacts
is of particular interest; so, the transfer integrals $t_{s,t}$ 
(and thence $\Gamma_{s,t}$, \emphj{cf.}, eqn~(\ref{eq-wbl})) are expected to be affected. 
}
\section{Conclusion}
\label{sec:conclusion}
We believe that the new theoretical results reported in the present paper offer relevant information for a better 
theoretical understanding of the microscopic processes occurring in molecular junctions away from equilibrium 
and useful hints for future experimental investigations.

We have identified experimental platforms allowing almost complete bias-controlled 
redox processes in molecular junctions. Information on the bias ($V$ and $V_G \sim \varepsilon_0$) 
dependent LUMO (in the specific case examined) 
occupancy $n$, the quantity on which we have focused our attention, can be obtained 
from bias dependencies of vibrational frequencies 
$\omega_{\nu}\left(V, \varepsilon_0\right) = 
\left[1 - n\left(V, \varepsilon_0\right)\right]\omega_{\nu}^{n} +
n\left(V, \varepsilon_0\right)\omega_{\nu}^{a} $
and Raman scattering intensities 
$A_{\nu}\left(V, \varepsilon_0\right) = 
\left[1 - n\left(V, \varepsilon_0\right)\right]A_{\nu}^{n} +
n\left(V, \varepsilon_0\right)A_{\nu}^{a} $
extracted from simultaneous SERS-transport measurements, which 
take values interpolating between the relevant (neutral $n$ and anionic $a$) 
charge species.\cite{Baldea:2014f} 
We have demonstrated that the experimental setup asymmetry plays an essential role
in achieving an almost perfect reduced state via bias tuning; efficient reduction is possible 
within some molecular electronics platforms, 
but definitely impossible under other platforms. 
We have shown that 
reduction can at most reach 50\% in two-terminal setups wherein a molecule 
is symmetrically coupled to electrodes (\figurename\ref{fig:ReductionAsymmetricPlatform}b). 
On the contrary, an almost complete redox process can be obtained in cases of highly 
asymmetric molecule-electrode couplings (\figurename\ref{fig:ReductionAsymmetricPlatform}b).
On this basis, an improved reduction can be expected in fullerene-based junctions if 
an STM platform (highly asymmetric coupling) is adopted instead of the electromigration 
platform (symmetric couplings) utilized in experiments.\cite{Natelson:14} 

Further, we found that an almost full reduction in two-terminal setups is accompanied
by current plateaus. Because such plateau effects have been observed in molecular junctions 
based on the zwitterionic molecule hexadecylquinolinium tricyanoquinodimethanide 
(\ce{C16H33Q-3CNQ}) 
\cite{Metzger:01} (seemingly the only known example), 
concurrent SERS-transport measurements \cite{Natelson:14,Baldea:2014f} on this system 
could be of interest to investigate bias-driven changes in molecular vibrational properties.

The fact that, unlike existing orbital gating measurements using ``dry'' platforms,\cite{Reed:09}
experiments resorting to electrolyte gating \cite{Tao:96,Alessandrini:06,Wandlowski:08} succeeded to 
reveal such a maximum renders the electrochemical three-terminal 
(EC-STM) platform a promising route 
in achieving an efficient reduction in molecular electronic devices,\cite{Baldea:2014f} 
also because it does not require high
biases hardly accessible experimentally and the coupling asymmetry is not critical 
(\emphj{cf.}~\figurename\ref{fig:i_n_vs_e0}b).
\section*{Acknowledgment}
Financial support provided by the 
Deu\-tsche For\-schungs\-ge\-mein\-schaft 
(grant BA 1799/2-1) is gratefully acknowledged.
%
\renewcommand\refname{Notes and references}
\providecommand*{\mcitethebibliography}{\thebibliography}
\csname @ifundefined\endcsname{endmcitethebibliography}
{\let\endmcitethebibliography\endthebibliography}{}

\end{document}